\documentclass{aa}

\def\dg{$^{\circ}$}

\usepackage{graphicx}
\usepackage{txfonts}
\begin{document}

   \title{Long photometric cycle and disk evolution in the $\beta$ Lyrae type binary OGLE-BLG-ECL-157529}

   \subtitle{}

   \author{R.E.\,Mennickent
          \inst{1}
          \and
          J.\,Garc\'es
          \inst{1}
          \and
          G.\,Djura\v{s}evi\'c
          \inst{2,3}
          \and
          P.\,Iwanek
          \inst{4}
          \and
          D.\,Schleicher
          \inst{1}
          \and
          R.\,Poleski
          \inst{4}
          \and
          I.\,Soszy{\'n}ski
          \inst{4}
          }

   \institute{Universidad de Concepci\'on, Departamento de Astronom\'{\i}a, Casilla 160-C, Concepci\'on, Chile\\
              \email{rmennick@udec.cl}
         \and
             Astronomical Observatory, Volgina 7, 11060 Belgrade 38, Serbia
                     \and
                     Issac Newton institute of Chile, Yugoslavia Branch, 11060, Belgrade, Serbia
                     \and 
                     Astronomical Observatory, University of Warsaw, Al. Ujazdowskie 4, PL-00-478 Warszawa, Poland
             }

   \date{Received XX XX, 2020; accepted XX XX, 2020}

 \titlerunning{The long photometric cycle of OGLE-BLG-ECL-157529}
\authorrunning{Mennickent et al.}

  \abstract
   {The subtype of hot algol semidetached binaries dubbed Double Periodic Variables (DPVs) are characterized  by a photometric cycle longer than the orbital one, whose nature has been related to a magnetic dynamo in the donor component controlling the mass transfer rate. }
   {We aim to understand the morphologic changes observed in the light curve of OGLE-BLG-ECL-157529 that are linked to the long cycle. In particular, we want to explain the changes in relative depth of  primary and secondary eclipses. 
}
   {We analyze $I$ and $V$-band OGLE photometric times series spanning 18.5 years and model the orbital light curve.}
   {We find that OGLE-BLG-ECL-157529 is a new eclipsing Galactic DPV of orbital period 24\fd8, and that its 
long cycle length decreases in amplitude and length during the time baseline. We show that the changes of the orbital light curve can be reproduced considering an accretion disk of variable thickness and radius, surrounding the hottest stellar component. Our models indicate changes in the temperatures of hot spot and bright spot during the long cycle, and also in the position of the bright spot. This, along with the changes in disk radius might indicate a variable mass transfer in this system.   }
   {}


   \maketitle
%

\section{Introduction}

Double Periodic Variables (DPVs) are a subset of Algol-type binary systems. They consist of a red giant that has filled its Roche lobe 
and transfers material through an optically thick accretion disk onto a B-type dwarf. Its main characteristic is that they have two 
photometric cycles: a short period P$_{\rm o}$, whose light curve shape is typical of the orbital modulation in eclipsing or ellipsoidal binaries and a long cycle period P$_{\rm l}$,  whose origin is still unknown. Both periodicities are 
related through the relationship P$_{\rm l}$ = 33 $\times$ P$_{\rm o}$, but the period ratio for a particular case can differ considerably from  the average   \citep{Mennickent2003, 2010AcA....60..179P, Pawlak2013,
Mennickent2016,  Mennickent2017}. In some systems, a decrease in the long period has been reported, as in the case of OGLE-LMC-DPV-065 \citep{2010AcA....60..179P, 2019MNRAS.487.4169M} and 
OGLE-LMC-DPV-056 \citep{Mennickent2008}.

A magnetic origin has been proposed as the cause of the long cycle \citep{2017A&A...602A.109S}.
The rapid rotation of an orbitally synchronized donor star, added to the convective motions would cause it to have a magnetic dynamo, which would change the equatorial stellar radius as indicated by the Applegate's mechanism \citep{1987ApJ...322L..99A, Applegate1992} which should modulate the mass transfer through the inner Lagrangian point. These changes could be observed as cyclic luminosity variations evidenced in the long cycle. This model, proposed by \citet{2017A&A...602A.109S},  predicts the correlation between the orbital and long periods and also the value of the long cycle length 
for particular systems with relatively good accuracy; for the seven studied binaries with orbital periods between 5 and 13 days, the maximum deviation between predicted and observed ratio is 30\%, with the average deviation of the order of 12\%. 
Actually, it has been shown that rapid rotation favors the operation of the Applegate's mechanism
\citep{2018A&A...615A..81N}.  

Other mechanisms have been identified as drivers for 
mass transfer changes in close binaries. These include direct modulation of mass transfer 
through the magnetic field of the donor star 
\citep{1989SSRv...50..311B, 2004MNRAS.352..416M}
and the effect of star spots or prominences \citep{1994ApJ...427..956L, 1996MNRAS.281..626S, 2004MNRAS.352..416M}.

Recently, Mg\,II, Fe\,I, Fe\,II, C\,I and Ti\,II  emission lines that are signatures of chromospherically active stars were detected in V\,393\,Scorpii, supporting the existence of magnetically active donors in DPVs \citep{2018PASP..130i4203M}.  In addition, magnetic fields have been inferred in $\beta$ Lyrae from the analysis of polarized light that could produce magnetically driven streams onto the accretion disk \citep{1982SvAL....8..126S, 2018CoSka..48..300S}.

Significant changes have been detected in the morphology of the light curve of OGLE-LMC-DPV-097 that are directly related to the
long cycle \citep{Garces2018}. These authors show that this kind of variability could be due to physical changes of the accretion disk, which would change its diameter and thickness cyclically according to the phase of the long cycle. If the long cycle reflects changes in the mass transfer rate, we might expect changes in disk properties
too, since the disk interacts with the gas stream.

\begin{figure*}
\scalebox{1}[1]{\includegraphics[angle=0,width=18cm]{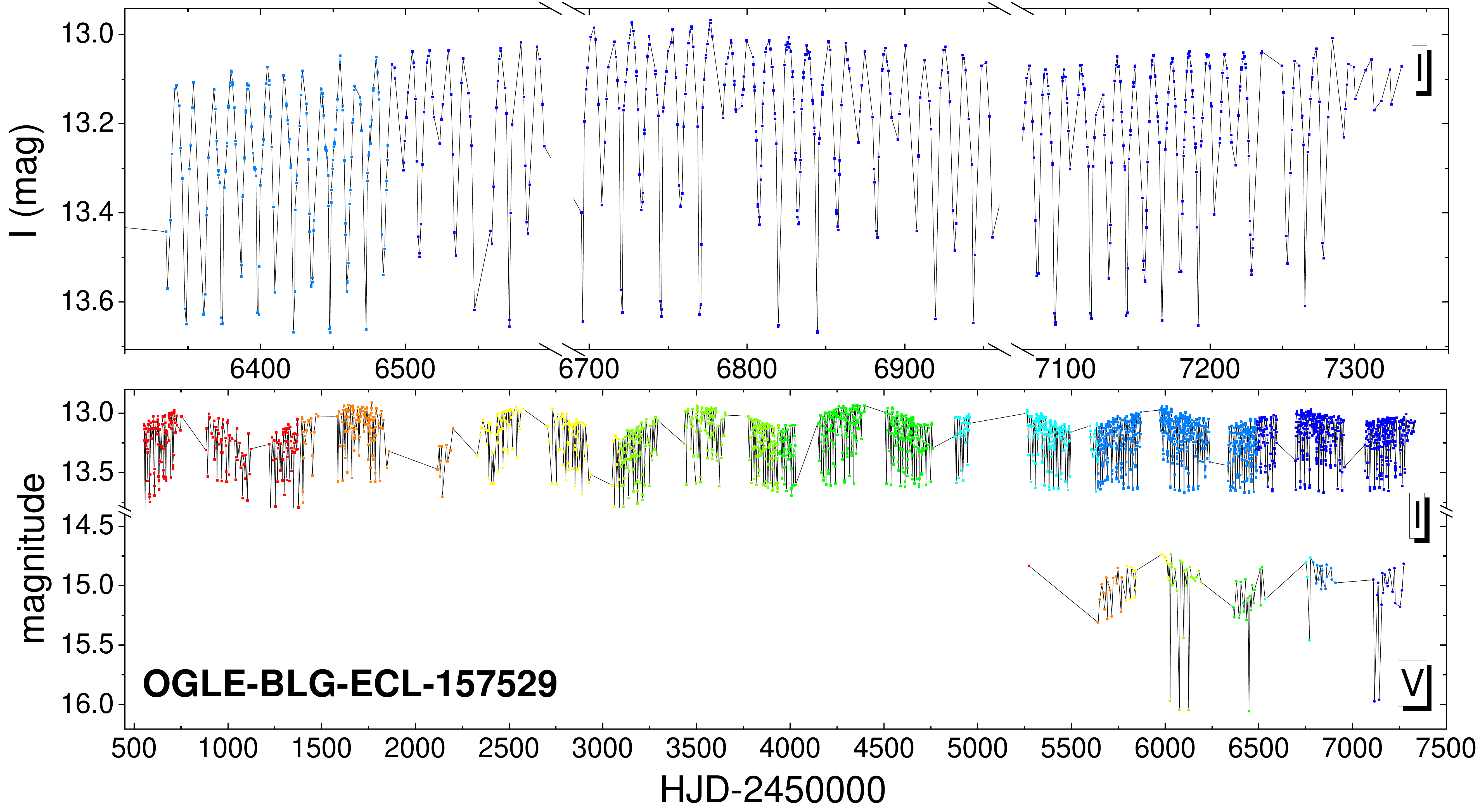}}
\caption{The OGLE $I$ and $V$-band light curves of OGLE-BLG-ECL-157529. Colors indicate different data ranges. }
\label{}
\end{figure*}

In this study we show the photometric analysis of a Galactic DPV binary system that presents a variable long cycle, viz.\, 
OGLE-BLG-ECL-157529 
($\alpha_{2000}$=17:53:08.33, $\delta_{2000}$=$-$32:46:27.0, $I$= 13.035 mag, $V$= 14.829 mag, \citet{Soszynski2016}). It is the first DPV reported in the direction of the Galactic bulge. The Gaia DR2 identification is 4043437999622564608 and its parallax is 0.331 $\pm$ 0.044 mas,  implying a distance of 3024 $\pm$ 406 pc \citep{2016A&A...595A...1G, 2018A&A...616A...1G}. This distance fits well
to that provided by \citet{2018AJ....156...58B}, viz.\,2815 pc 
with lower limit at 2487 pc and upper limit at 3240 pc.
The above indicates that the systems is not  a member of the Galactic  bulge, but a foreground star.

We used good time coverage in OGLE data to study changes in the morphology of the light curve. It occurred that they are, similarly to OGLE-LMC-DPV-097, related to the long cycle.
Our motivation is to investigate the physical phenomena that give rise to 
 the  peculiar changes  observed in its orbital light curve, in order to understand the cause of these changes. 
This object, together with other DPVs  that show linked orbital and long cycle variability, could be fundamental  in a study aimed to understand the DPV 
phenomenon and test the hypothesis of the magnetic dynamo.  
A preliminary report of this investigation was presented by \citet{2019CoSka..49..355G}.

\section{Photometric Data}

This object is included in the catalogue of eclipsing binaries in the Galactic bulge presented by \cite{Soszynski2016}. 
The photometric time series analyzed in this study consists of 2606 $I$-band data points  taken from the following data bases (Fig.\,1): 
OGLE-II  \citep{2005AcA....55...43S}\footnote{http://ogledb.astrouw.edu.pl/$\sim$ogle/photdb/} and OGLE-III/IV\footnote{ 
OGLE-III/IV data kindly provided by the OGLE team.}. The OGLE-IV project is described by \citet{2015AcA....65....1U}.
In addition, the OGLE-IV data base provides 118  additional $V$-band measurements.
The whole dataset, summarized in Table\,1, spans a time interval of  6781 d, i.e. 18.5 yr.

\begin{figure}
\scalebox{1}[1]{\includegraphics[angle=0,width=8.5cm]{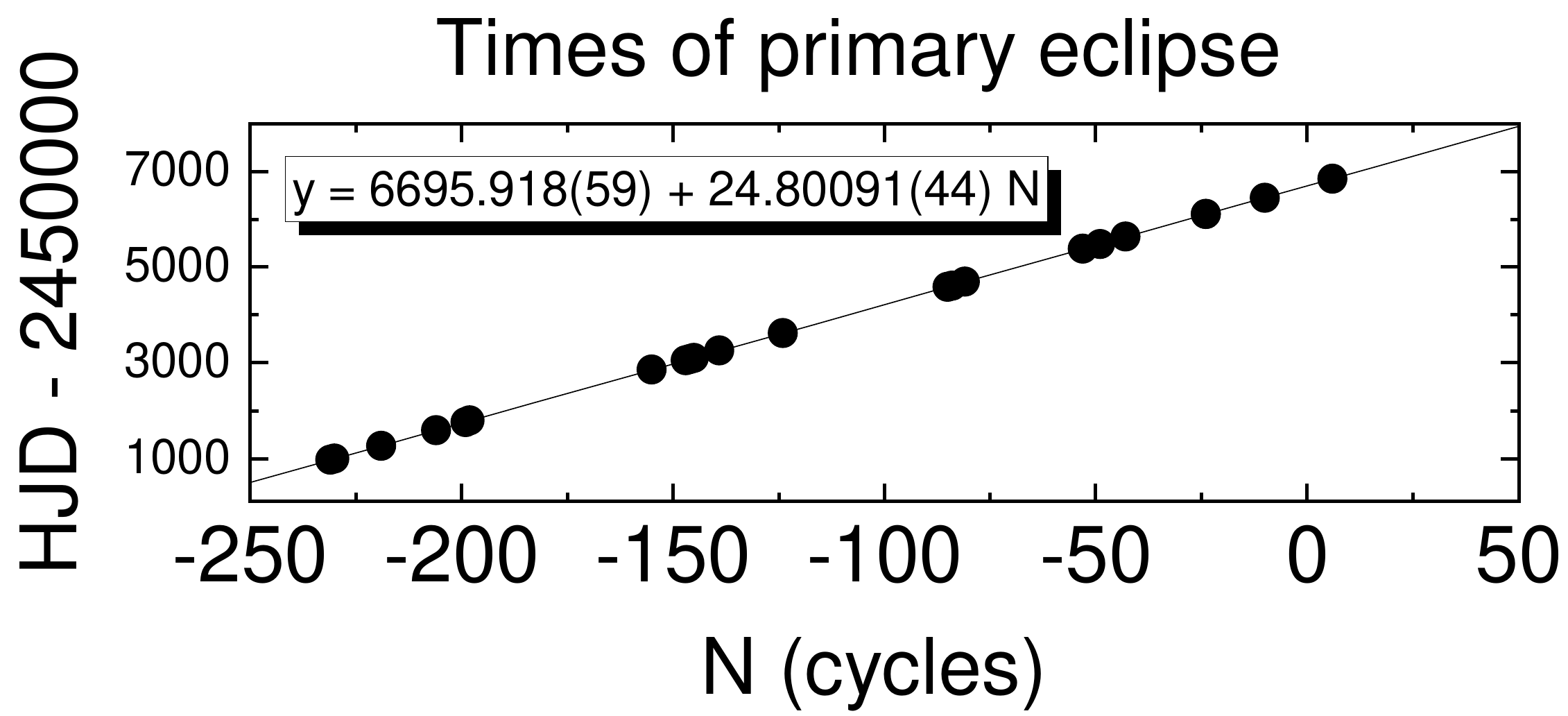}}
\caption{Times of primary eclipse provide an improved value for the orbital period $24\fd80091 \pm \rm 0\fd00044$. }
\label{}
\end{figure}

\begin{figure*}
\scalebox{1}[1]{\includegraphics[angle=0,width=18cm]{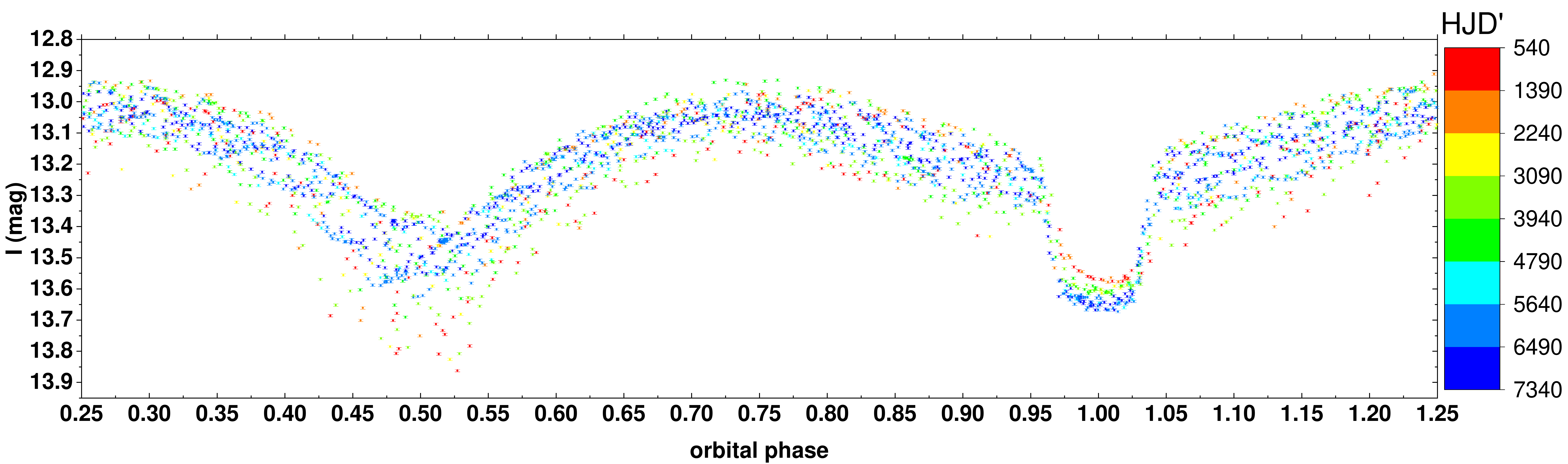}}
\caption{The OGLE $I$-band light curve of OGLE-BLG-ECL-157529 phased with the  ephemeris given by Eq.\,1. Colors label data shown in Fig.\,1. }
\label{}
\end{figure*}

\begin{figure}
\scalebox{1}[1]{\includegraphics[angle=0,width=8.5cm]{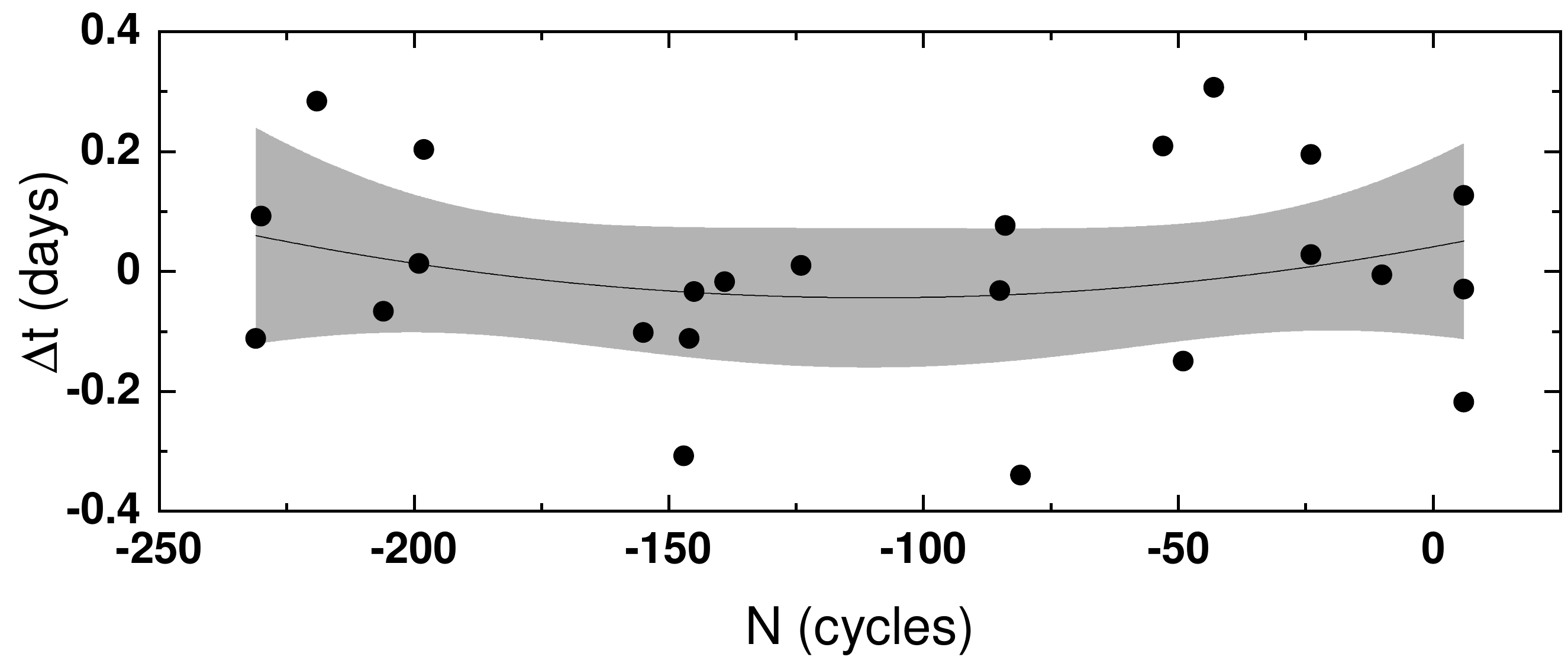}}
\caption{Observed minus calculated times for primary eclipses using the orbital period given by the  ephemeris of Eq.\,(1) and the best fit. Dashed area shows the 95\% confidence band.}
\label{}
\end{figure}

\begin{figure}
\scalebox{1}[1]{\includegraphics[angle=0,width=8.5cm]{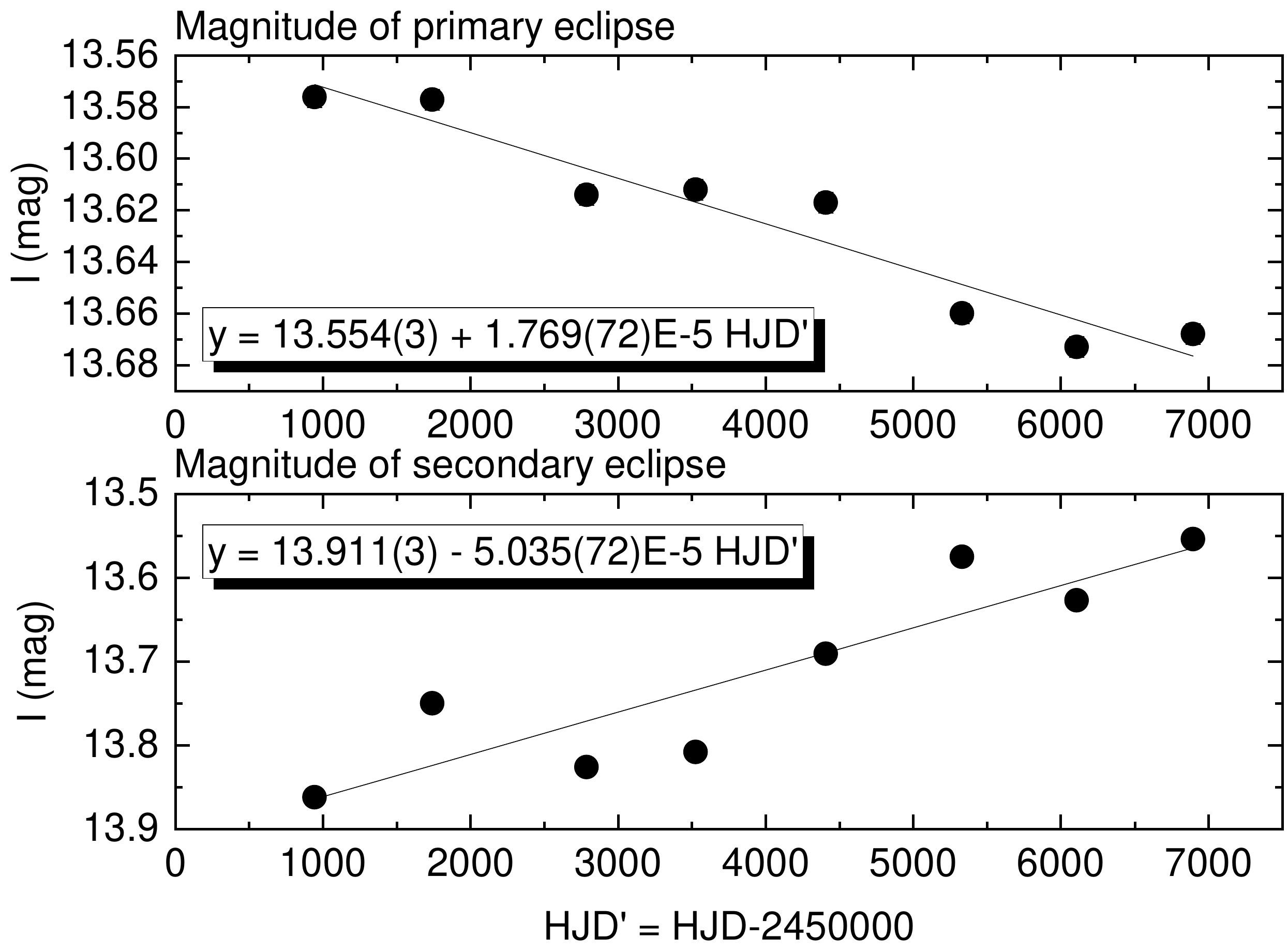}}
\caption{Eclipse magnitudes as a function of time. The lowest point of every colored data segment in the eclipses of Fig.\,2 is plotted.}
\label{}
\end{figure}

\begin{figure}
\scalebox{1.1}[1.1]{\includegraphics[angle=0,width=8.5cm]{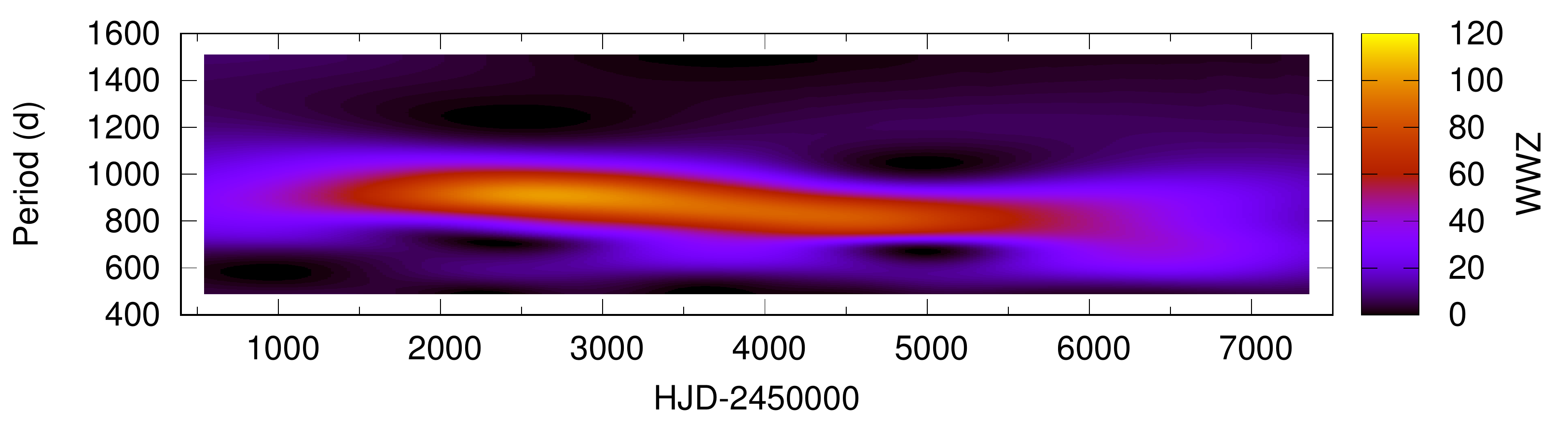}}
\caption{The WWZ 2D power spectrum as a function of period and time.}
\label{}
\end{figure}

\begin{table}
\centering
\caption{Summary of photometric observations. The number of measurements, starting and ending times, and average magnitude are given. 
HJD' = HJD-2450000. The uncertainty of a single measurement varies between  4 and 6 mmag. }
\label{tab:example_table}
\begin{tabular}{lcccc} 
\hline
\small
band/Data-Base &  N      & $HJD'_{start}$      &  $HJD'_{end}$   & Mag      \\
\hline
I / OGLE-II    & 346     &   551.77073   &  1858.52147  &   13.213 \\
I / OGLE-III     & 795     &   2117.76494  &  4955.73490  &   13.221  \\
I / OGLE-IV      & 1465    &   5261.84891  &  7332.50545  &   13.213  \\
V/ OGLE-IV &118 & 5274.88451&7273.60682 &15.038 \\
\hline
\end{tabular}
\end{table}

\begin{figure*}
\scalebox{1}[1]{\includegraphics[angle=0,width=18cm]{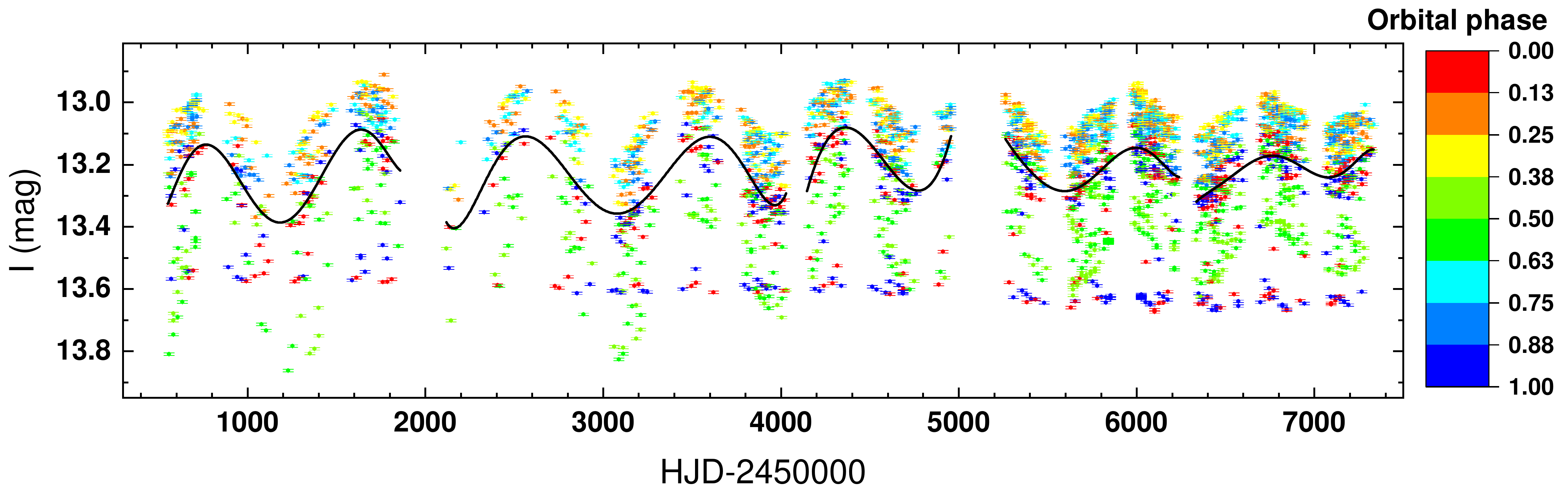}}
\caption{The $I$-band light curve showing the polynomial fits to the long cycle. Colors represents the orbital phases according to the  ephemeris given by Eq.\,1. Note the evolution of the secondary eclipse (green) and the primary eclipse (red and blue).}
\label{}
\end{figure*}

\begin{figure}
\scalebox{1}[1]{\includegraphics[angle=0,width=8.5cm]{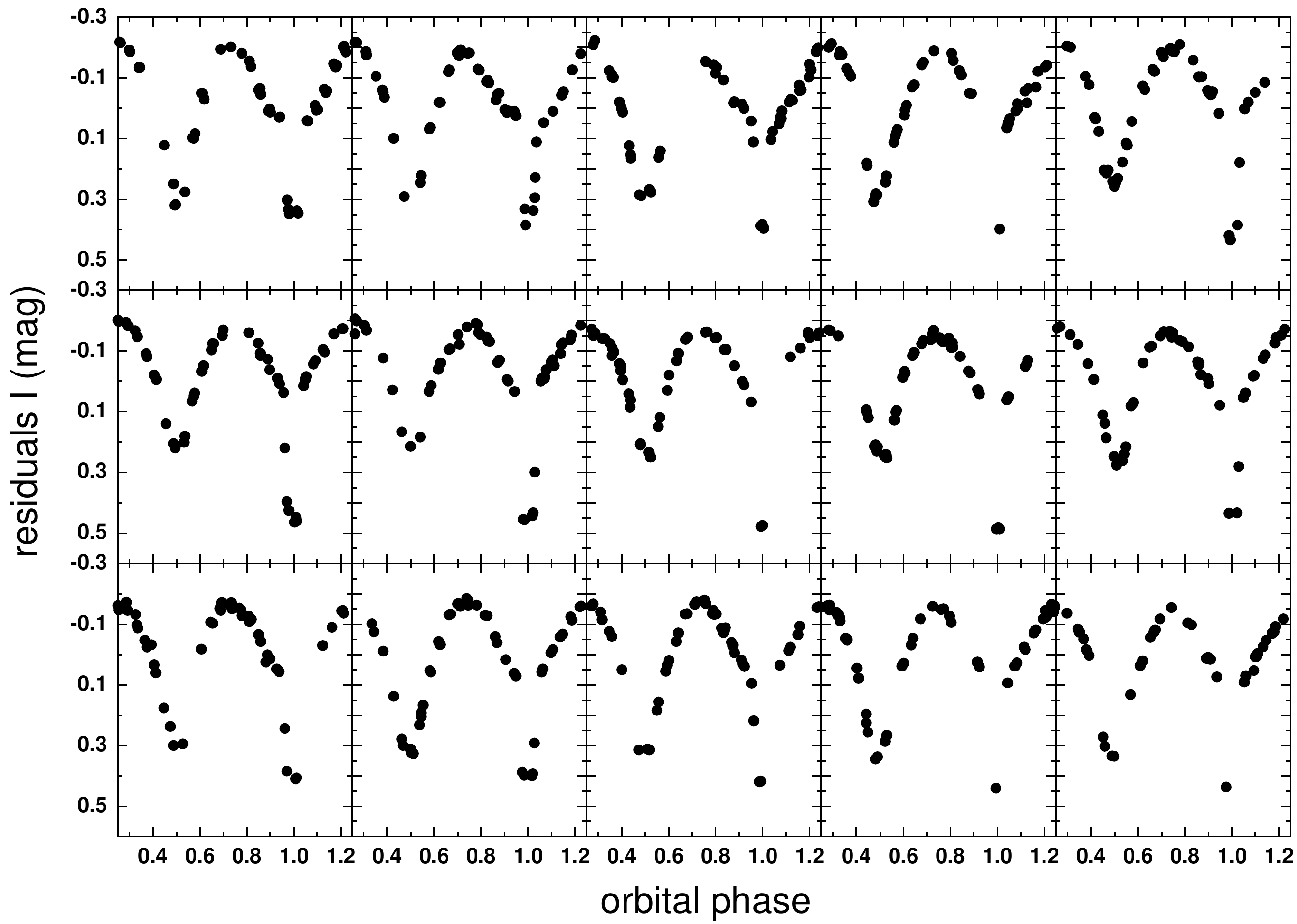}}
\caption{Orbital light curves constructed with residuals of the fits shown in fig.\,7 for data spanning the time interval 
(HJD-245000) between 6334.9 and 7332.5. Every panel shows 50 consecutive data points.
Time goes from left to  right from upper left to lower right panel.}
\label{}
\end{figure}

\begin{figure}
\scalebox{1}[1]{\includegraphics[angle=0,width=8.5cm]{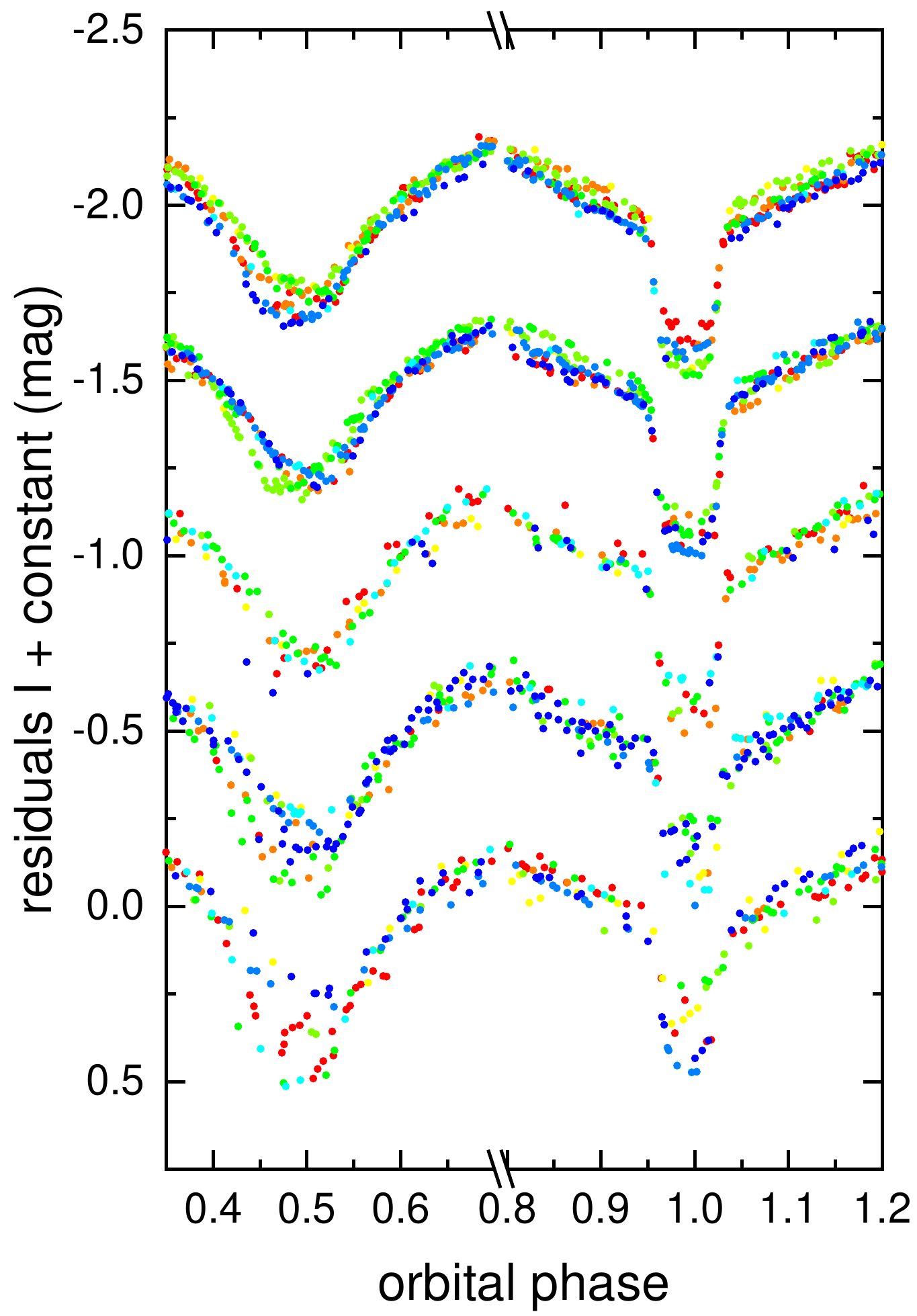}}
\caption{Orbital light curves constructed with the residuals of the fits shown in fig.\,7.  Data are grouped for the first fit (lower) until the fifth fit residuals (upper).
To illustrate changes in the shape of the light curves, data are marked with a color map following the HJD; red for the first observations and blue for the last observations of every segment.}
\label{}
\end{figure}

\begin{figure}
\scalebox{1}[1]{\includegraphics[angle=0,width=8.5cm]{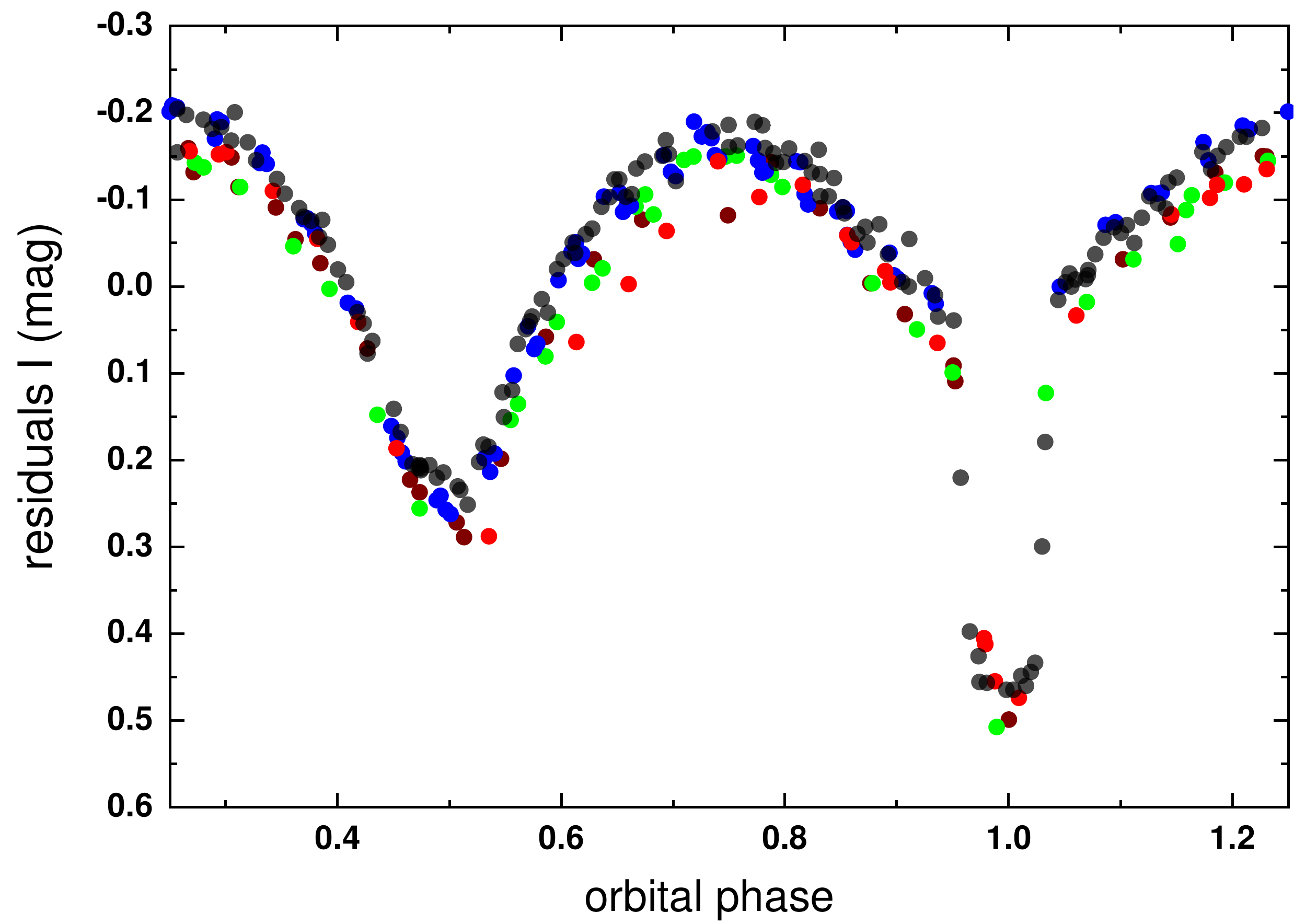}}
\scalebox{1}[1]{\includegraphics[angle=0,width=8.5cm]{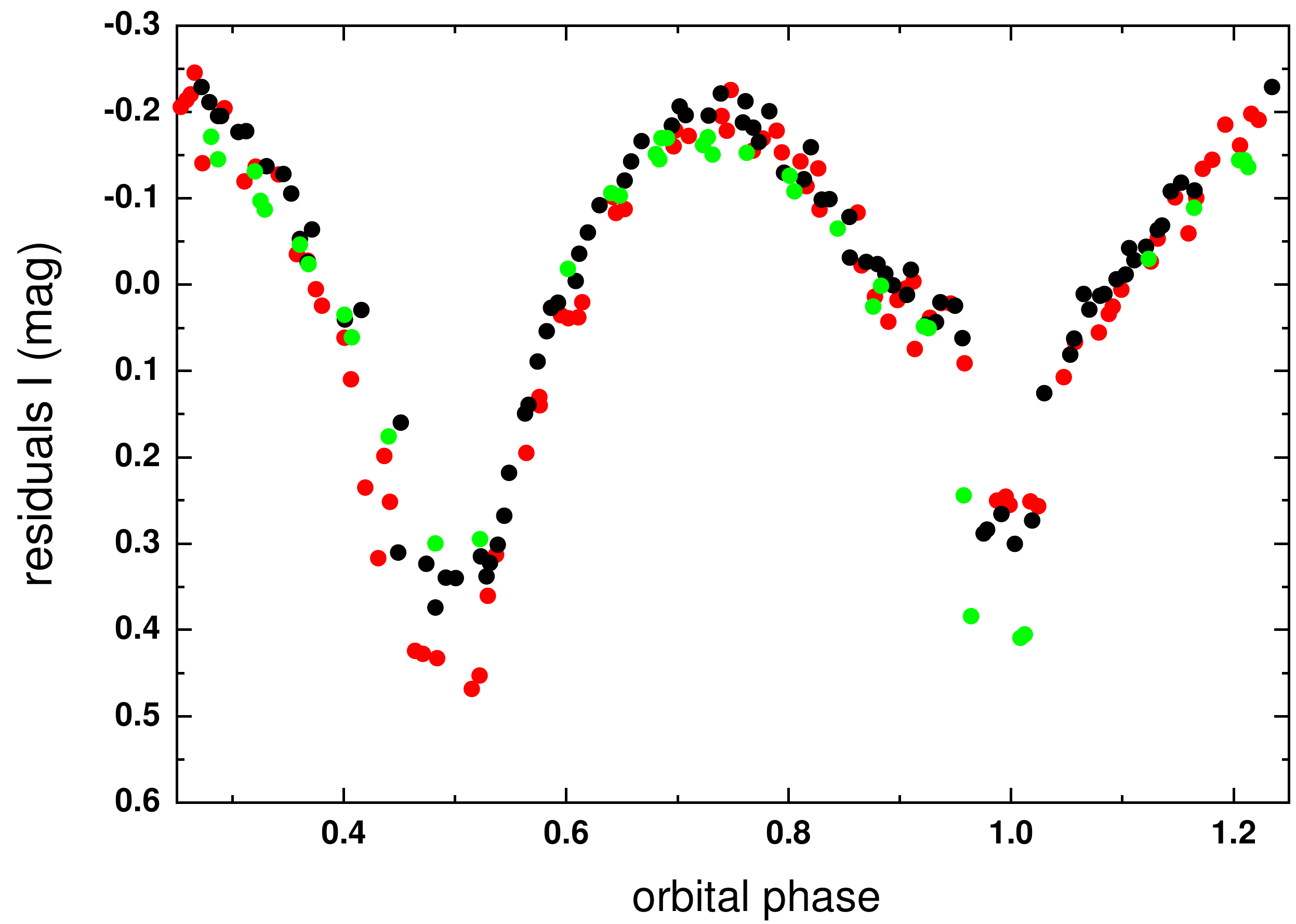}}
\caption{Orbital light curves constructed with the residuals of the fits shown in fig.\,5.  Data are grouped for times of maxima (upper) and minima (lower) of the long cycle.
The ranges of HJD - 2450000 are: (Upper) brown; 3569.8-3629.6, green; 4329.6-4393.5, blue; 5866.5-6023.8, red; 1608.8-1667.8 and black; 6572.6-6816.8,
(Lower) red; 3049.9-3205.6,
black; 3881.7-4028.5 and green; 7064.9-7110.9.  }
\label{}
\end{figure}

\section{Light curve analysis}

The light curve shows alternate and periodic minima that are typical of an eclipsing binary system, but also a long cycle in time scale of several hundreds of days (Fig.\,1).
The system is fainter in the $V$-band than in the $I$-band and the eclipses are deeper in $V$.  

We used the generalized Lomb-Scargle (GLS) periodogram to obtain the orbital frequency in our data. This algorithm was introduced by 
\citet{2009A&A...496..577Z}  and uses the principle of the \citet{1976Ap&SS..39..447L} \& \citet{1982ApJ...263..835S} periodograms with some modifications, such as the addition of a displacement in the adjustment of the fit function and the consideration of measurement errors. Compared with the classical periodogram, it gives us more accurate frequencies and a better determination of the amplitudes. We obtained 
P$_{\rm o} =  24\fd7992 \pm \rm 0\fd0024.$ 

With the above period we phased the $I$-band light curve. This allowed us to select the deeper data points in the primary eclipse, corresponding in this system to the usually more deeper eclipse, for every data segment that is colored in Fig.\,1. These data points were
analyzed with the cycle-number technique for linear-ephemerides \citep{2005ASPC..335....3S}  and the following improved  ephemeris for the primary eclipse was derived (Fig.\,2): 
\small
\begin{eqnarray}
\rm HJD&=&\rm (245\,6695\fd918 \pm \rm 0\fd059) + \rm (24\fd80091 \pm \rm 0\fd00044)\, E
\end{eqnarray}
\normalsize

\noindent
where $N$ is the cycle number. This period was used to phase the light curve shown in Fig.\,3. From this figure we infer the following: (1) at some epochs the secondary eclipse becomes deeper than the primary eclipse, (2)  the secondary eclipse has larger variability (scatter) than the primary eclipse, (3) the regularity of time strings reflected in colored data suggests that the non-orbital variability occurs mostly in time scales of hundred of days, (4) while the secondary eclipse changes its depth as well as its width, the primary one make it primarily in depth, and  with less amplitude than the secondary eclipse,  (5) the long cycle, revealed by the scattered data, becomes more evident outside the primary eclipse, suggesting that the light source becomes at least partially eclipsed during primary eclipse, (6) there is a tendency for the primary eclipse in being deeper with time, and (7) the orbital period seems to be very stable during the observing period.

We tested the last statement making an analysis of the times of primary eclipse with the "period-change equation"  \citep{1980A&A....82..172P}. This method assumes a constant rate of change of the
period and compares  predicted times with observed times in a similar way that  the O-C diagram described by  \citet{2005ASPC..335....3S} does it. As eclipse timings we choose the fainter points in the colored data strings  shown in Fig.\,3.  
We find a period rate of change of dP$_{\rm o}$/dt = (1.41 $\pm$ 1.38) $\times$ 10$^{-5}$ d d$^{-1}$,  i.e. the period is constant for all practical purposes (Fig.\,4). 

We find that the eclipses depth changes linearly with time and in opposite way on the 18.5-year interval; the total change of the secondary eclipse depth, viz.\, about 0.35 mag, is larger than the total change of the primary eclipse depth, viz.\, about 0.10 mag (Fig.\,5).

The long cycle is revealed in the Weighted Wavelet Z transform (WWZ) as defined by \citet{1996AJ....112.1709F}. The WWZ works similarly as the Lomb-Scargle periodogram providing information about the periods of the signal and the time associated to those periods. It is very suitable for the analysis of non- stationary signals and has advantages in the analysis of time-frequency local characteristics. The WWZ shows the long period decreasing from around 900 days to around 800 days through the time baseline (Fig.\,6). 

We fit data segments with order-6 polynomials in order to represent the long cycle. These fits show that the long cycle decreases its period and amplitude during the time baseline (Fig.\,7).
It is clear from this figure the continuous smooth decrease of the brightness of the primary eclipse (lower red and blue points, orbital phases around 0 or 1) along with the much fainter secondary eclipses occurring at some epochs (green very low points, orbital phases around 0.5). It is also clear how the secondary eclipse becomes brighter as time goes by. 

We analyzed the orbital light curves after removing the long cycle. This was done using the residuals of the fits shown in Fig.\,7. Examples of  "cleaned" orbital light curves are  shown in Fig.\,8. These show
changes in the shape of the orbital light curve which are more evident during the first observing epochs (Fig.\,9). In particular, we notice the changes in the shape of the eclipses and the ingress and egress light curve branches. The primary eclipse is not always deeper than the secondary one, this is specially evident in the first observing epochs. We also notice changes in the shape of the light curve during epochs of long cycle maximum and minimum (Fig.\,10); during the minima of the long cycle occurs the reversals of the eclipse depths, something already visible in Fig.\,7 at the lower parts of the fits. 

In order to understand these changes we investigate in the next section the nature of the star facing the observer  during the primary eclipse. This is
usually named donor star in a $\beta$-Lyr type binary, because it transfer mass onto the "gainer" star through a gas stream flowing from its inner Lagrangian point.

\subsection{On the system reddening, distance and donor star}

Since we do not have spectroscopic data for our object, we must rest in the photometry to further understand the system. 
Using the OGLE-IV 
photometry and the disentangled orbital light curve, we obtain a $V-I$ = 2.42 mag at the primary eclipse. 
The location of this DPV in the direction of the Galactic bulge causes that the light from this object is strongly
reddened by unevenly distributed interstellar matter. 

This object is located in the OGLE field BLG535, which is distant from the Galactic plane by slightly over 3 degrees. In this direction dust is distributed highly non-uniformly and reddening changes on a small angular scale. \citet{2013ApJ...769...88N} have estimated reddening using red clump stars located in the Galactic bulge, so reddening measurements from Nataf's map are the upper limits toward analyzed directions, and they should not be used straightforward for the objects located definitely closer than the Bulge. 
In  Fig.\,11 we show  a color-magnitude diagram (CMD) towards the direction of the OGLE-BLG-ECL-157529. We use stars located in the radius
of about 2.5 arcmin around the position of the analyzed object. In the CMD we mark our target, the red clump (which has mean position:
(($V-I$), $I$) = (2.33 mag, 16.11 mag) and the reddening vector.
We use the OGLE Extinction Calculator\footnote{http://ogle.astrouw.edu.pl/cgi-ogle/getext.py} that uses the Nataf's extinction maps, and obtain
limit for the color excess E($V-I$) = 1.266 mag using the nearest method and E($V-I$) = 1.248 mag using the
natural neighbor method.

As previously mentioned, the
Gaia parallax of 0.33065 $\pm$ 0.04436 mas implies a distance of $d$ = 3024 $\pm$ 406 pc. This means that the object is in the Galactic disk and not in the Bulge. 
Using the Recio-Blanco model for the Galactic disk \citep{2014A&A...567A...5R} along with the Gaia distance we obtain  E($V-I$) = 0.945 mag implying a dereddened color index of $(V-I)_{0}$= 1.475 mag at the  primary eclipse. 
In the previous step we used as input the uncorrected color excess obtained with the natural neighbor method.  To complement our research on reddening, we used also 
the Python package named "mwdust"\footnote{https://github.com/jobovy/mwdust} that uses several existing extinction maps in the Milky Way \citep{2016ApJ...818..130B}. Using this package, we obtain extinction at a given direction and distance. 
For the distance 3.02 kpc we have obtained 
extinction in  $V$- and $I$-band A$_V$ = 2.288 mag and A$_I$ = 1.256 mag, what gives E($V-I$) = 1.033 mag roughly confirming the earlier calculation.

The corrected color in  primary eclipse obtained with the Recio-Blanco model \citep{2014A&A...567A...5R}, viz.\, $(V-I)_{0}$= 1.47 mag, corresponds to a K2-type star with 
$\rm T_{c}$ = 4400\,K \citep{Drilling2000} (Fig.\,12). In the following we will assume that this is the  effective temperature of the donor star. The relative stability of the primary eclipse might support this conjecture.

\begin{figure}
\begin{center}
\begin{tabular}{ll}
\includegraphics[width=0.90\linewidth]{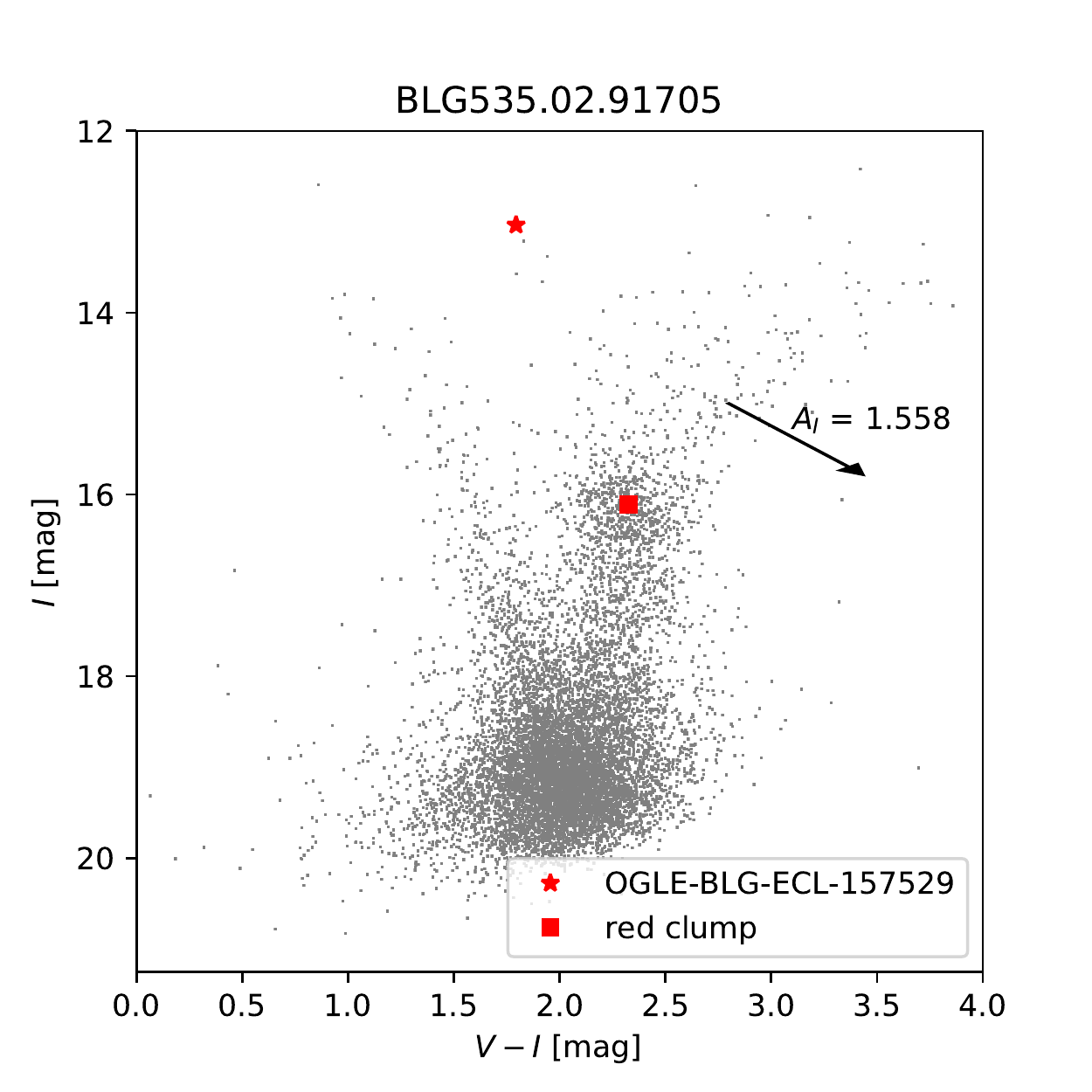}
\end{tabular}
\end{center}
\caption{color-magnitude diagram for stars surrounding OGLE-BLG-ECL-157529. The reddening vector, extinction value and magnitudes of the red clump are from \citet{2013ApJ...769...88N}. More details can be found in Section 3.1.}
\label{D1}
\end{figure}

\begin{figure}
\scalebox{1}[1]{\includegraphics[angle=0,width=8.5cm]{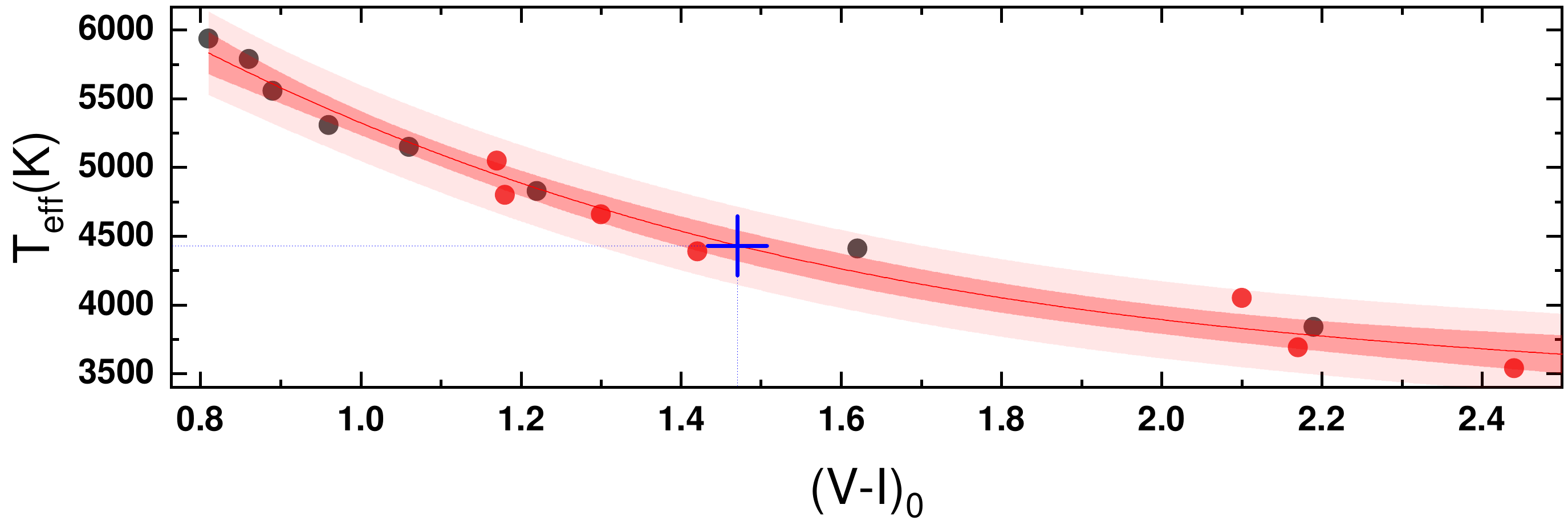}}
\caption{Temperature versus color for dwarfs (black dots) and giants (red dots) from \citet{Drilling2000} and the best 3$^{th}$ order polynomial fit. 95\% confidence and prediction bands are shown by dashed and  dashed-light areas, respectively. The position of OGLE-BLG-ECL-157529 during primary eclipse is also shown. }
\label{}
\end{figure}

\begin{figure}
\scalebox{1}[1]{\includegraphics[angle=0,width=8.5cm]{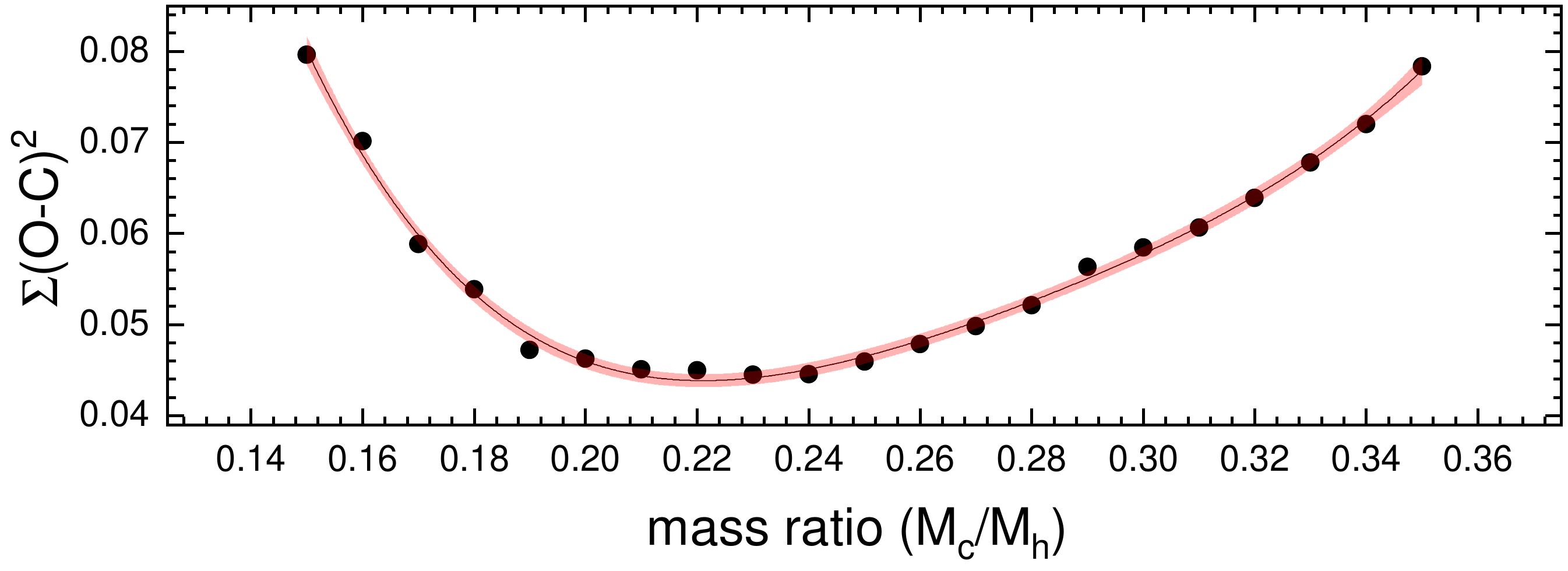}}
\caption{The parameter $\Sigma$ (O-C)$^{2}$ for the fits done to the light curve with data obtained at the maximum of the long cycle, as a function of mass ratio. 
The best 4$^{th}$ order polynomial fit is shown along with the 95\% confidence band.}
\label{}
\end{figure}

\section{Discussion}

\subsection{Overall qualitative interpretation}

The overall behavior of the light curve, especially the behavior of the secondary eclipse, can be interpreted in the terms of circumstellar material in the system.
The changing shape of the secondary eclipse cannot be due to occultation by a star, but by a variable structure such as an accretion disk, surrounding the more massive star and formed by  mass transfer from the Roche lobe overflow of the less massive star. This configuration is typical of semidetached Algols of the $\beta$ Lyrae type. What is not typical and it is unique for this system is the presence of a large amplitude long photometric cycle and the remarkable changes observed in the relative depths of the primary and  secondary eclipses.  

From the analysis shown in the previous section we might infer the existence of a large disk during the first observing epochs, but that decreases its 
size when time goes by. This should explain why the secondary eclipse is deeper on the beginning. This long-term behavior of the disk is reflected in the secondary eclipse magnitude (Fig.\,5). When the disk is vertically larger (regarding the orbital plane), it occults a larger fraction of the donor and gainer  and the eclipse becomes deeper. However, the deeper secondary eclipses occurs only during long-cycle minima, and this should indicate that the disk size is also modulated by the long cycle, but in a shorter time scale compared with the aforementioned decade-length secular tendency. 
 
We also notice that the disk attains maximum size when the long cycle has larger amplitude and its cycle length is longer, i.e. at the beginning of the time series. 
We suggest that changes in mass transfer controls the disk size and eventually the brightness of the extra source responsible of the long cycle. This extra source might be a hot spot wind as suggested for the DPV system V\,393 Scorpii
\citep{2012MNRAS.427..607M} and also for $\beta$ Lyrae \citep{1996A&A...312..879H}. Since we don't detect changes in the orbital period, the mass transfer in this system should be rather small.  

\subsection{Light curve model}

In this subsection we model the light curve in order to test the picture of  the variable disk given above. For that we use a theoretical code that solves the inverse problem and that considers a system consistent of a  donor star and a  gainer star surrounded by an accretion disk, that is both optically and geometrically thick
\citep{1992Ap&SS.196..267D, 1992Ap&SS.197.17D, 1996Ap&SS.240..317D}. The disk temperature is the same that the gainer in its inner edge and decreases with a radial profile described by an exponent $a_T$. The model includes a hot spot and a bright spot in the disk, following evidence found in previous observations of algols \citep{2004AN....325..229R}. These active regions influence the shape of the light curve during the ingress and egress of the eclipses. The model has been described in detail in several papers, so we remit the interested reader to \citet{2013MNRAS.432..799M, bib:paper6, 2019MNRAS.487.4169M}.
We assume a donor temperature  $\rm T_{c}$ = 4400\,K, as derived in the previous section. We also assume synchronous rotation for the  donor, as expected for a close binary that rapidly synchronize stellar spins with the orbit due to tidal forces.  On the other hand, the gainer might have been spun-up
to a high rotation due to nearly tangentially infalling material \citep{1981A&A...102...17P}, hence we have assumed critical rotation for this star. We find that our model 
is practically the same for synchronous and critical gainer.

We use the q-search method, usually applied to over-contact or semi-detached binaries when no spectroscopic data are available \citep{2005Ap&SS.296..221T}. We find  convergent solutions for a range of mass ratios $q= M_{c}/M_{h}$, where the subindexes "c" and "h" refer to the cool and hot stars. A fit with a fourth order polynomial to the corresponding $\Sigma$ $(O-C)^{2}$  values, yields an optimal mass ratio $q$  = 0.22 (Fig.\,13). Previous studies show $\overline{q}$ = 0.23 $\pm$ 0.05 (standard deviation) for the few studied DPVs \citep{Mennickent2016}. For the q-search we used data obtained at maximum (Fig.\,10 upper), when the disk influence should be lower, as discussed in the previous section.

Next we model the light curve at data ranges representative of the maximum and minimum of the long cycle. We choose those data ranges that best represent the observed differences between maximum and minimum, where the inversion of the eclipses is more evident. The parameters of the fits are shown in 
Table\,2 and a comparison of our theoretical models and observations is given in Fig.\,14. While the models reproduce relatively well the general appearance of the light curves, the fit at minimum has more scatter than at maximum. It is posible that at minimum the emissivity sources are more complex and our simple model of disk plus spots cannot reproduce so well the overall system emissivity. 

\subsection{Discussion of the light curve models}

Our model indicates a system seen at angle 85\dg and a stellar separation of 65 ${\rm R_{\odot}}$. The stellar temperatures are 4400 $K$ and 14.000 $K$. The stellar masses are 4.83 and 1.06 ${\cal M_{\odot}}$ and surface gravities log\,g = 3.86 and 2.02. The stellar radii are  4.5 and 16.6 ${\rm R_{\odot}}$. The gainer
has a surface temperature indicating a spectral type B\,6 \citep{1988BAICz..39..329H}, consistent with the B-type spectral types found in DPV gainers. Although we have justified the use of the fixed parameters $T_c$ and $q$, the stellar parameters might be refined when spectroscopic data come to be available.  

Significant differences are observed in disk properties when passing from maximum to minimum of the long cycle: 
the radius increases from  27 to 33 ${\rm R_{\odot}}$, while the temperature at the outer edge remains around  3000-3500 $K$. The vertical thickness at the outer edges increases from 3 to 11 ${\rm R_{\odot}}$, which explains the inversion of the eclipse depths.  The average disk radius of  30  ${\rm R_{\odot}}$ means ${\rm R_{d}/a}$ $\approx$ 0.47, i.e. the disk outer border is just below the tidal radius for the given mass ratio \citep{1977ApJ...216..822P, 1995CAS....28.....W}. This opens the possibility that at minimum the outer disk, especially optically thin regions that are not tested with our light curve model, becomes influenced by tidal forces. The light contribution of the larger disk explains the shallower primary eclipse observed at the first epochs (red points in Fig.\,3).

The bright spot is located roughly at the expected region where the stream impacts the disk, 47\dg\ apart from the lines joining the stellar centers in the direction of the orbital motion. This position does not change between maximum and minimum. On the contrary, the position of the bright spot changes from  66\dg\ to  125\dg\
from maximum to minimum, as measured in the opposite direction of the orbital  motion. At maximum the hot spot temperature is  48\% higher than the surrounding disk, 
at minimum is only  13\% higher, in principle consistent with much kinetic energy released in the stream-disk impact region in a smaller disk. At maximum the bright spot temperature is  37\% higher than the surrounding disk and 8\% at minimum. 

We notice that changes in disk parameters, especially disk radius and temperatures of the active regions, might indicate variable mass transfer in the system, as required in the dynamo model proposed by \citet{2017A&A...602A.109S}.

\begin{figure}
\scalebox{1}[1]{\includegraphics[angle=0,width=8.5cm]{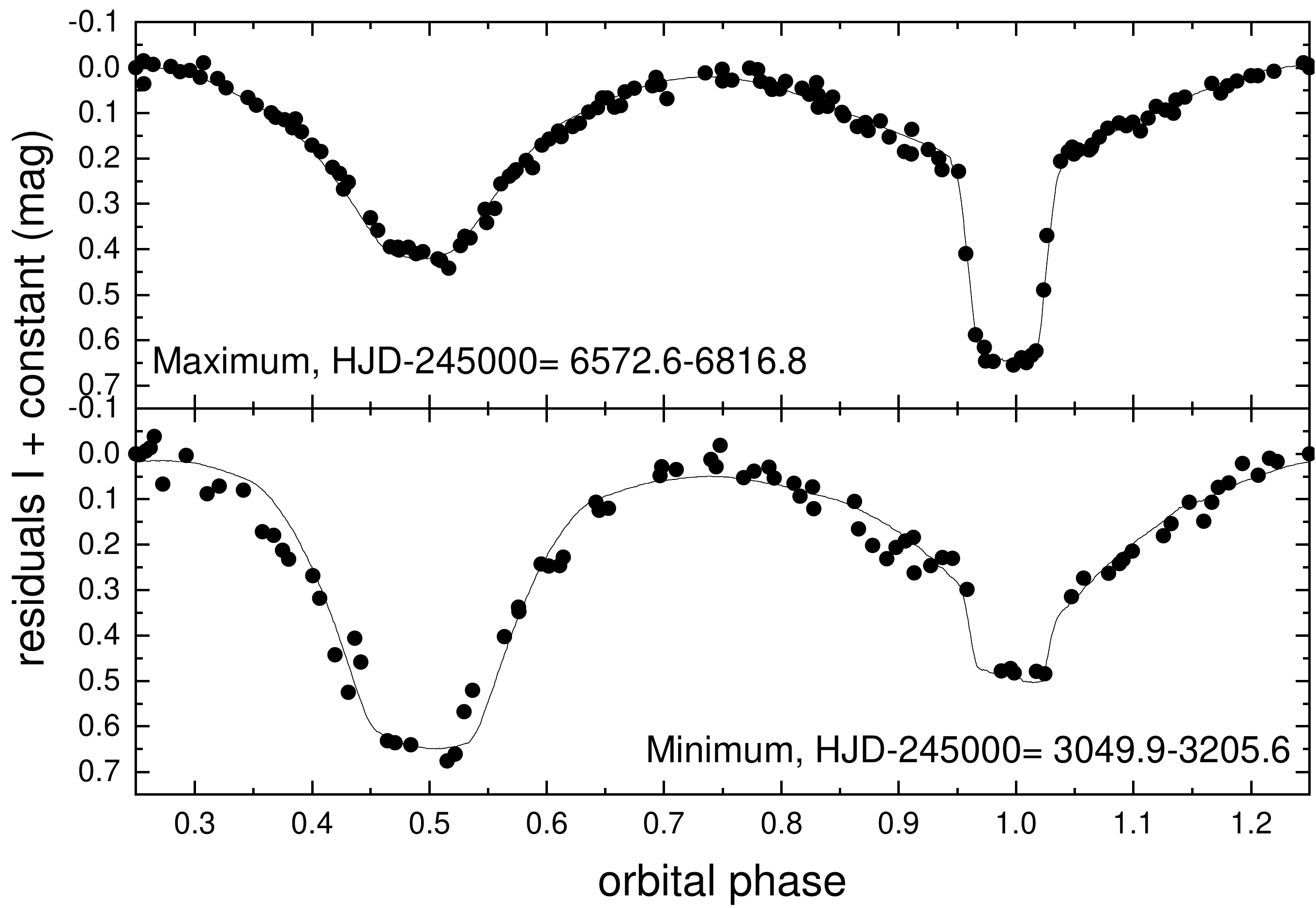}}
\caption{Orbital light curves at long cycle maximum and minimum along with the fits provided by the models described in Table\,2. }
\label{}
\end{figure}

\section{Conclusion}

\begin{itemize}

\item We find that the eclipsing binary OGLE-BLG-ECL-157529 is a Double Periodic Variable characterized by an orbital period of 24\fd80091 $\pm$  0\fd00044 and a  long cycle length decreasing in length and amplitude during 18.5 years of observations. 

\item The overall light curves can be understood in terms of a variable accretion disk. The disk is larger and thicker at long cycle minimum and this effect is more pronounced when the long cycle has larger amplitude and longer cycle length.  

\item Our models indicate changes in the temperatures of hot spot and bright spot during the long cycle, and also in the position of the bright spot. This, along with the changes in disk radius might indicate variable mass transfer in this system.

\end{itemize}

\begin{acknowledgements}
 We thanks the anonymous referee who contributed to improve the first version of this manuscript. REM and JG acknowledge support by VRID-Enlace 216.016.002-1.0, BASAL Centro de Astrof{\'{i}}sica y Tecnolog{\'{i}}as Afines (CATA) PFB--06/2007 and FONDECYT 1190621. DS and REM thank FONDECYT 1201280 and DS thanks FONDECYT 1161247.  
JG acknowledges ANID project 21202285 and members of stellar variability group 
(S.V.G. UdeC) for useful discussions about this work. GD  acknowledges the financial support of the Ministry of Education, Science and Technological Development
of the Republic of Serbia through the contract No 451-03-68/2020/14/20002. 
The OGLE project has received funding
from the Polish National Science Centre grant MAESTRO no. 2014/14/A/ST9/00121.
This work  has made use of data from the European Space Agency (ESA) mission
{\it Gaia} (https://www.cosmos.esa.int/gaia), processed by the {\it Gaia}
Data Processing and Analysis Consortium (DPAC,
https://www.cosmos.esa.int/web/gaia/dpac/consortium). Funding for the DPAC
has been provided by national institutions, in particular the institutions
participating in the {\it Gaia} Multilateral Agreement.
 
\end{acknowledgements}

\begin{table*}

\caption{
Results of the analysis of the light curves shown in  Fig.\,14. 
The parameters are obtained by solving the inverse problem for the Roche model with
an accretion disk around the more massive (hotter) gainer in
 critical non-synchronous rotation regime \citep{1992Ap&SS.196..267D, 1992Ap&SS.197.17D, 1996Ap&SS.240..317D}.}

 \label{TabOGLE-max-min}
      \[
        \begin{array}{llllllll}
            \hline
{\rm Quantity} & {\rm OGLE-max}& {\rm Quantity}& {\rm OGLE-max} & {\rm Quantity} &{\rm OGLE-min} & {\rm Quantity}&{\rm OGLE-min}  \\
            \noalign{\smallskip}
            \hline
            \noalign{\smallskip}
   n                               & 142             & \cal M_{\rm_h} {[\cal M_{\odot}]} & 4.83  \pm 0.3 &    n                               & 85              & \cal M_{\rm_h} {[\cal M_{\odot}]} & 4.83  \pm 0.3 \\
{\rm \Sigma(O-C)^2}                & 0.0386          & \cal M_{\rm_c} {[\cal M_{\odot}]} & 1.06  \pm 0.2 &  {\rm \Sigma(O-C)^2}               & 0.1404          & \cal M_{\rm_c} {[\cal M_{\odot}]} & 1.06  \pm 0.2\\
{\rm \sigma_{rms}}                 & 0.0166          & \cal R_{\rm_h} {\rm [R_{\odot}]}  & 4.48  \pm 0.2 &  {\rm \sigma_{rms}}                & 0.0409          & \cal R_{\rm_h} {\rm [R_{\odot}]}  & 4.48  \pm 0.2 \\
   i {\rm [^{\circ}]}              & 85.4  \pm 0.2   & \cal R_{\rm_c} {\rm [R_{\odot}]}  & 16.6  \pm 0.2 &     i {\rm [^{\circ}]}             & 85.5  \pm 0.3   & \cal R_{\rm_c} {\rm [R_{\odot}]}  & 16.6  \pm 0.2\\
{\rm F_d}                          & 0.816 \pm 0.03  & {\rm log} \ g_{\rm_h}             & 3.82  \pm 0.1 &  {\rm F_d}                         & 0.99  \pm 0.03  & {\rm log} \ g_{\rm_h}             & 3.82  \pm 0.1\\
{\rm T_d} [{\rm K}]                & 2970  \pm 200   & {\rm log} \ g_{\rm_c}             & 2.02  \pm 0.1 &  {\rm T_d} [{\rm K}]               & 3560  \pm 300   & {\rm log} \ g_{\rm_c}             & 2.02  \pm 0.1\\
{\rm d_e} [a_{\rm orb}]            & 0.045 \pm 0.005 & M^{\rm h}_{\rm bol}               &-2.24  \pm 0.2 &  {\rm d_e} [a_{\rm orb}]           & 0.163 \pm 0.006 & M^{\rm h}_{\rm bol}               &-2.31  \pm 0.2\\
{\rm d_c} [a_{\rm orb}]            & 0.073 \pm 0.005 & M^{\rm c}_{\rm bol}               &-0.13  \pm 0.1 &  {\rm d_c} [a_{\rm orb}]           & 0.032 \pm 0.007 & M^{\rm c}_{\rm bol}               &-0.13  \pm 0.1\\
{\rm a_T}                          & 7.7   \pm 0.3   & a_{\rm orb}  {\rm [R_{\odot}]}    & 64.6  \pm 0.3 &  {\rm a_T}                         & 7.4   \pm 0.4   & a_{\rm orb}  {\rm [R_{\odot}]}    & 64.6  \pm 0.3\\
{\rm f_h}                          & 36.5  \pm 0.5   & \cal{R}_{\rm d} {\rm [R_{\odot}]} & 27.4  \pm 0.3 &  {\rm f_h}                         & 36.4  \pm 0.5   & \cal{R}_{\rm d} {\rm [R_{\odot}]} & 33.2  \pm 0.3\\
{\rm F_h}                          & 1.000           & \rm{d_e}  {\rm [R_{\odot}]}       & 2.9   \pm 0.2 &  {\rm F_h}                         & 1.000           & \rm{d_e}  {\rm [R_{\odot}]}       & 10.5  \pm 0.3\\
{\rm T_h} [{\rm K}]                & 14080 \pm 500   & \rm{d_c}  {\rm [R_{\odot}]}       & 4.7   \pm 0.2 &  {\rm T_h} [{\rm K}]               & 13990 \pm 500   & \rm{d_c}  {\rm [R_{\odot}]}       & 2.0   \pm 0.3\\
{\rm A_{hs}=T_{hs}/T_d}            & 1.48  \pm 0.05  &                                                   && {\rm A_{hs}=T_{hs}/T_d}           & 1.13  \pm 0.03  &                                                  \\
{\rm \theta_{hs}}{\rm [^{\circ}]}  & 18.8  \pm 2.5   &                                                   && {\rm \theta_{hs}}{\rm [^{\circ}]} & 14.9  \pm 2.6   &                                                  \\
{\rm \lambda_{hs}}{\rm [^{\circ}]} & 313.0 \pm 6.0   &                                                   && {\rm \lambda_{hs}}{\rm [^{\circ}]}& 314.6 \pm 7.0   &                                                  \\
{\rm \theta_{rad}}{\rm [^{\circ}]} & 7.7   \pm 7.0   &                                                   && {\rm \theta_{rad}}{\rm [^{\circ}]}& 12.5  \pm 7.0   &                                                  \\
{\rm A_{bs}=T_{bs}/T_d}            & 1.37  \pm 0.04  &                                                   && {\rm A_{bs}=T_{bs}/T_d}           & 1.08  \pm 0.03  &                                                  \\
{\rm \theta_{bs}}{\rm [^{\circ}]}  & 47.2  \pm 8.0   &                                                   && {\rm \theta_{bs}}{\rm [^{\circ}]} & 21.9  \pm 3.0   &                                                  \\
{\rm \lambda_{bs}}{\rm [^{\circ}]} & 66.2  \pm 9.0   &                                                   && {\rm \lambda_{bs}}{\rm [^{\circ}]}& 125.4 \pm 9.0   &                                                  \\
{\Omega_{\rm h}}                   & 17.848\pm 0.02  &                                                   && {\Omega_{\rm h}}                  &17.838 \pm 0.03  &                                                  \\
{\Omega_{\rm c}}                   & 2.282 \pm 0.02  &                                                   && {\Omega_{\rm c}}                  & 2.282 \pm 0.02  &                                                  \\
            \noalign{\smallskip}
            \hline
         \end{array}
      \]

\small FIXED PARAMETERS: $q={\cal M}_{\rm c}/{\cal M}_{\rm
h}=0.22$ - mass ratio of the components, ${\rm T_c=4400 K}$  -
temperature of the less massive (cooler) donor, ${\rm F_c}=1.0$ -
filling factor for the critical Roche lobe of the donor,
$f{\rm _{c}}=1.00$ - non-synchronous rotation coefficients
of the donor, ${\rm F_h}=R_h/R_{zc}$ - filling factor for the
critical non-synchronous lobe of the hotter, more massive gainer (ratio of the
stellar polar radius to the critical Roche lobe radius along z-axis for
a star in critical non-synchronous rotation regime), ${\rm \beta_h=0.25}$, 
${\rm \beta_c=0.08}$ - gravity-darkening coefficients of the components, ${\rm A_h=1.0}$,
${\rm A_c=0.5}$  - albedo coefficients of the components.
NOTE: $n$ - number of observations, ${\rm
\Sigma (O-C)^2}$ - final sum of squares of residuals between
observed (LCO) and synthetic (LCC) light-curves, ${\rm
\sigma_{rms}}$ - root-mean-square of the residuals, $i$ - orbit
inclination (in arc degrees), ${\rm F_d=R_d/R_{yc}}$ - disk
dimension factor (the ratio of the disk radius to the critical Roche
lobe radius along y-axis), ${\rm T_d}$ - disk-edge temperature,
$\rm{d_e}$, $\rm{d_c}$,  - disk thicknesses (at the edge and at
the center of the disk, respectively) in the units of the distance
between the components, $a_{\rm T}$ - disk temperature
distribution coefficient, $f{\rm _h}$ - non-synchronous rotation coefficient
of the more massive gainer (in the critical non-synchronous rotation regime),
${\rm T_h}$ - temperature of the gainer, ${\rm
A_{hs,bs}=T_{hs,bs}/T_d}$ - hot spot temperature
coefficients, ${\rm \theta_{hs,bs}}$ and ${\rm \lambda_{hs,bs}}$ -
spot angular dimension and longitude (in arc degrees), ${\rm
\theta_{rad}}$ - angle between the line perpendicular to the local
disk edge surface and the direction of the hot-spot maximum
radiation, ${\Omega_{\rm h,c}}$ - dimensionless surface potentials
of the hotter gainer and cooler donor, $\cal M_{\rm_{h,c}} {[\cal
M_{\odot}]}$, $\cal R_{\rm_{h,c}} {\rm [R_{\odot}]}$ - stellar
masses and mean radii of stars in solar units, ${\rm log} \
g_{\rm_{h,c}}$ - logarithm (base 10) of the system components
effective gravity, $M^{\rm {h,c}}_{\rm bol}$ - absolute stellar
bolometric magnitudes, $a_{\rm orb}$ ${\rm [R_{\odot}]}$,
$\cal{R}_{\rm d} {\rm [R_{\odot}]}$, $\rm{d_e} {\rm [R_{\odot}]}$,
$\rm{d_c} {\rm [R_{\odot}]}$ - orbital semi-major axis, disk
radius and disk thicknesses at its edge and center, respectively,
given in solar units.

\end{table*}

\end{document}